\documentclass[aps,prl,showpacs,reprint,floats,epsfig,pdflatex,superscriptaddress,twocolumn]{revtex4}
\usepackage{epsfig}
\usepackage{amsmath}
\usepackage{amsfonts}
\usepackage{graphicx}
\usepackage{amssymb}
\usepackage{amsbsy}
\usepackage{epstopdf}
\usepackage{subfigure}
\usepackage{tabularx}
\usepackage{color}
\graphicspath{{Pictures/}} 
\usepackage{subfig}
\usepackage{ mathrsfs }
\usepackage{physics}
\usepackage[justification=justified,singlelinecheck=false]{caption}

\begin{document}

\title {Magnetic coherent population trapping in a single ion}
\author{S.~Das}
\author{P.~Liu}
\affiliation{Centre for Quantum Technologies, National University Singapore, Singapore 117543}
\author{B.~Gr\'{e}maud}
\author{M.~Mukherjee}
\affiliation{Centre for Quantum Technologies, National University Singapore, Singapore 117543}
\affiliation{Department of Physics, National University Singapore, Singapore 117551}
\affiliation{MajuLab, CNRS-UNS-NUS-NTU International Joint Research Unit, UMI 3654, Singapore}

\date{\today}

\begin{abstract}
Magnetically induced coherent population trapping has been studied in a single trapped laser cooled ion. The magnetic field dependent narrow spectral feature is found to be an useful tool in determining the null point of magnetic field at the ion position. In particular, we use a double lambda scheme that allows us to measure the null magnetic field point limited by the detector shot noise. We analyzed the system theoretically and found certain long lived bright states as the dark state is generated under steady state condition.
\end{abstract}


\maketitle


\section{introduction}
Light matter interaction plays a vital role to understand the atomic structure,
to manipulate the quantum states of an atom and for application external field
sensing \cite{Woo97,Dut16,Bud07,Tsi09,Muk04,Lud15}. In the case of a two state atom interacting with light field, the interaction is well characterized by the Rabi formalism. However this formalism is valid only in cases where the two atomic states are well separated from any other states of the atom and the interacting light has a narrow linewidth as compared to the separation. The off-resonant coupling however grows larger with intensity of the light field \cite{Stalnaker}. This leads to loss of coherence of the atomic state. To circumvent the problem, different strategies are adopted, namely by using an extra light field to compensate the off-resonant coupling or to perform quantum manipulation in a state which is decoupled from the light field itself. In the field of quantum computation the second approach is presently been explored by employing Stimulated Raman Rapid Adiabatic Process (STIRAP) where the qubit is formed by the dark states and the coupling is obtained by far-off-resonance excitation to a bright state~\cite{Berk04,Zhou17}. As the name suggests the dark states are completely decoupled from the addressing light (no absorption or emission occurs) and hence to observe them one needs to apply a second field to couple the dark states to a fluorescing state or bright state. Since a dark state is weakly coupled to other states also means that such a state is generally immune to external perturbations as well and hence they are useful resource for sensing~\cite{Bel07}. In sensing applications, the darkness of these states leads to phenomena like coherent population trapping~(CPT) and electromagnetically induced transparency~(EIT). These are very similar phenomena as one obtains in a three level lambda-type system with two ground states and an excited state addressed by two coherent light fields. The light fields under certain condition, produce a coherent superposition state between the two ground states leading to decoupling of the light field from the upper state of the lambda system. This leads on one hand to transparency (no absorption) of the probe light while population trapping of the atom in a dark state formed by the two ground states \cite{Arimondo} on the other hand. These phenomena have been studied extensively both in theory as well as in experiment \cite{Kubo}. In particular, the CPT in atomic ensemble leads to high sensitive magnetic field probes limited in spatial resolution due to the requirement of an ensemble of atoms to increase the signal-to-noise-ratio \cite{Groeger}. In zero magnetic field, it is still possible to obtain a CPT, but signal is independent of the laser frequency and hence frequency is not an usable resource for magnetic field probe. Only very recently, this has been explored in terms of developing narrow clock transition and the phenomena is re-named as magnetically induced CPT or MCPT \cite{peter}. Theoretically, it has been studied in the context of laser cooling of ions in an ion trap where MCPT is detrimental to cooling as the ions get trapped in the dark state and hence remain out of the cooling cycle\cite{Berk04}. To the best of our knowledge, MCPT on a double lambda system has never been studied either theoretically or experimentally. Here we study on a single barium ion the MCPT in a double lambda configuration.  This particular system is interesting because: (i) it allows to study the phenomena of MCPT at a single atom level; (ii) the double lambda system, as we will show later, makes the zero magnetic field measurement background free; (iii) this system provides an insight on the possibility of manipulating multiple MCPTs formed within the system which may have relevance to dark-state quantum computation or sensing. In order to be usable, the dark state should efficiently couple to a bright state in order to read out the final quantum state.\\

In this article, we first discuss the experimental method followed by the experimental setup,
procedure and results. We conclude by a discussion on some of the
possible applications of this single ion double lambda MCPT system.

\section{Method}
The basic principle of operation is based on the level crossing
spectroscopy \cite{Franken}.
\begin{figure}[htbp]
        \centering
        \includegraphics[width=.45\textwidth]
        {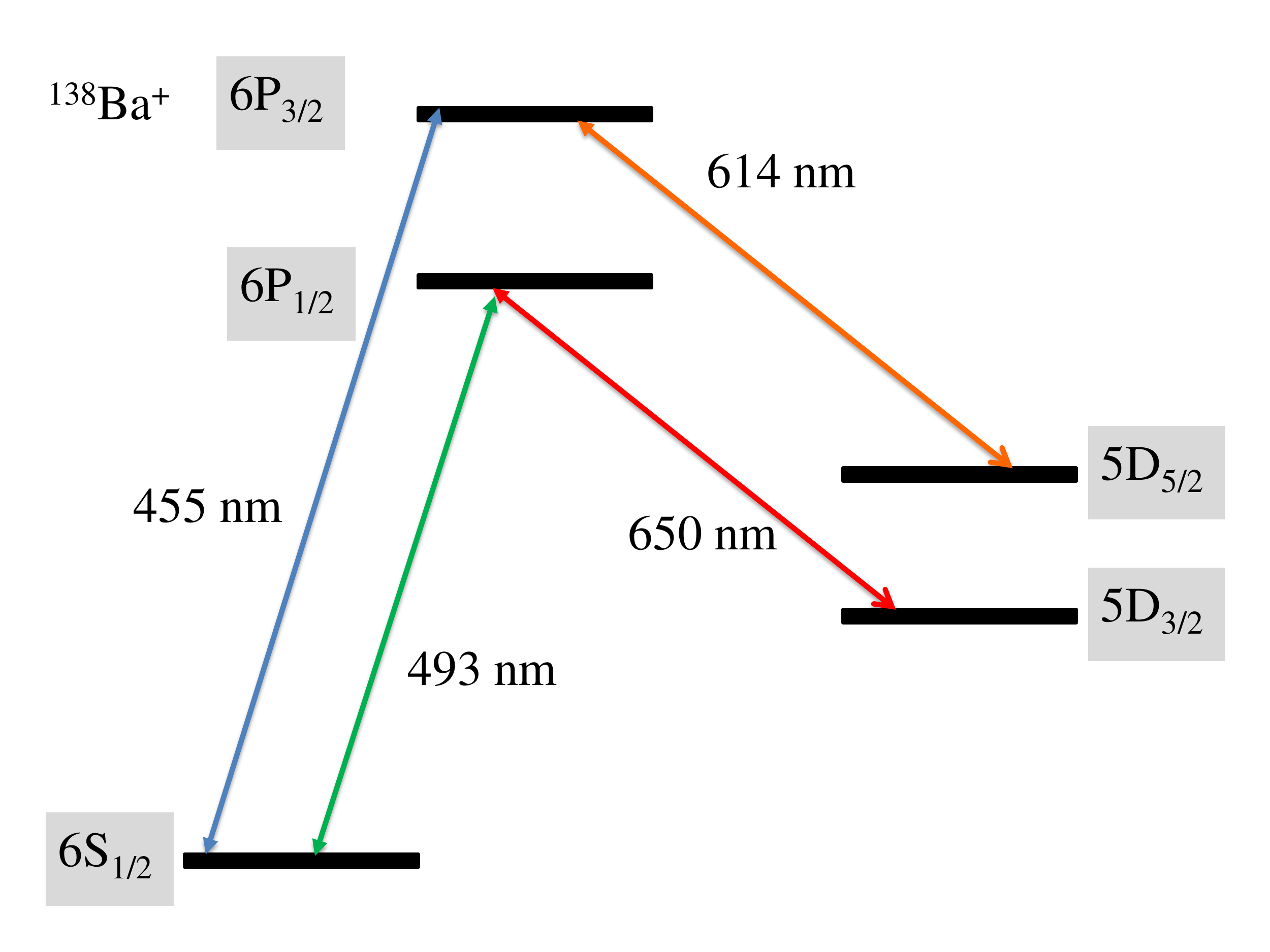}
    \caption[$\textrm{Ba}^+$ level structure.]{Relevant energy levels of $^{138}\textrm{Ba}^+$ along with the wavelengths of transitions. It shows two lambda type energy levels.}
    \label{fig:level_diagram}
\end{figure}

For simplicity we will assume, a five level system with three ground (two meta-stable and one ground) states and two excited state forming a double $\Lambda$-type level
scheme as shown in Fig.~(\ref{fig:level_diagram}). We will be referring the S-P$_{3/2}$-D as the upper lambda and the S-P$_{1/2}$-D as the lower lambda system. In our case the ion is
a single $^{138}$Ba$^+$, however as long as the energy level is
similar to any alkaline atom, this scheme can be implemented. The ion
is laser cooled via the S$_{1/2}$-P$_{1/2}$ transition at $493$~nm. The fluorescence at $493$nm or $455$nm is observed by continuously exciting the S$_{1/2}$-P$_{3/2}$ transition
at $455$nm simultaneously. Lasers at $614~$nm and $650~$nm are simultaneously applied to pump out the population trapped in D$_{5/2}$ and D$_{3/2}$ respectively. The rate of observed fluorescence depends on frequency detuning and intensities of the lasers as well as the
strength of the external magnetic field. Here, we are interested in the dependence of the fluorescence counts on the magnetic field strength, near to the zero field. For the special case of zero magnetic field, the Zeeman sub-levels of the electronic D-levels are
degenerate. This magnetically induced degeneracy leads to the origin of dark states within the
fine-structure manifold of the D-levels. These states due to their difference in multiplicity are not coupled to the excited states P$_{1/2}$ and P$_{3/2}$ levels independent of the laser detuning, polarization, and intensity. In electronic structure of barium ion each D-level supports two dark states, which makes the present scheme extremely robust to any experimental imperfections. Therefore one obtains a narrow resonance like feature about the zero magnetic field which is also
known as the coherent population trapping~\cite{Ali05}. As observed in the experiment, the width of this resonance is mainly governed by the saturation parameters of
the re-pumpers provided, the $455~$nm laser intensity is below saturation. Fundamentally, the width is limited by the combined line width of the applied lasers. \\

While CPT is widely used in magnetometry with gas cells, MCPT suffers from velocity dependent shifts and hence not much useful~\cite{Wyna99}. On the contrary a Doppler cooled single ion does not suffer from velocity dependent shifts but the signal-to-noise ratio is low. Therefore we employ here a double lambda scheme to make the zero field fluorescence measurement background free. Thus the spectroscopic feature studied here is fundamentally different as compared to MCPT in a gas cell. \\

\begin{figure}[htbp]
        \centerline{
        \includegraphics[width=.48\textwidth]
        {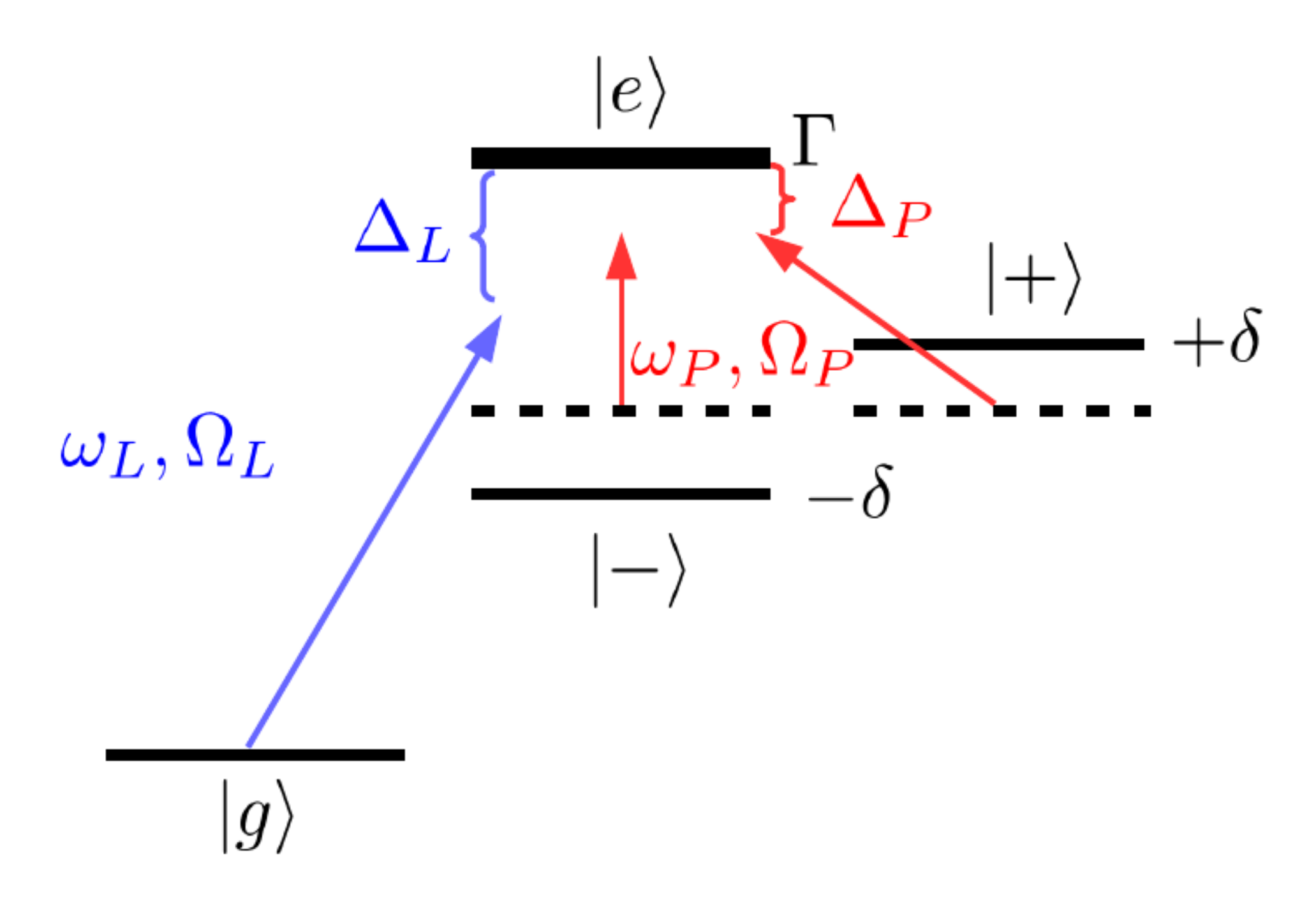}}
    \caption{(Color Online)A schematic diagram of a simple model that exhibits the same feature as the experimental scheme.  As long as the
two states have the same energy, i.e. $\delta=0$, they form a perfect two-photon resonant $\Lambda$ system. In the situation where the two lasers have
different detunings ($\Delta_L\ne\Delta_P$), the four level system exhibits one single dark state $(|+\rangle-|-\rangle)/\sqrt{2}$, i.e. fully independent
of the different laser parameters. In the stationary regime, the atom is shelved in this dark state and, therefore, is not scattering any photons. }
    \label{fig:simple_model}
\end{figure}

From a theoretical point of view, the properties of the fluorescence light are obtained from the stationary solution of the full optical Bloch equations, i.e. including all $18$ levels,
assuming a diagonal density matrix in the S$_{1/2}$ states for the initial state. The external motion is not taken into consideration as it plays insignificant role for the temperature of the ion used in the experiment. In order to extract the ultimate sensitivity of the scheme, the rate of change of the fluorescence intensity as a function of the applied magnetic field is necessary. The derivative of the fluorescence $\partial I/\partial B$ and its slope $\partial^2I/\partial B^2$ are obtained from the Taylor expansion of this stationary solution with respect to the magnetic field, i.e. amounting to doing a perturbation expansion to first and second order. As explained above, the crucial ingredient of the experimental scheme is the existence of a dark state due to a lambda type coupling. A toy model exhibiting the same properties is shown in Fig.~\ref{fig:simple_model}: the ground state $|g\rangle$ is coupled to the excited state $|e\rangle$ by the first laser (detuning $\Delta_L$), while the two metastable states $|\pm\rangle$ are coupled by the same laser (detuning $\Delta_P$). As long as the two states have the same energy, i.e. $\delta=0$ corresponding to the dashed lines, they form a perfect two-photon resonant $\Lambda$ system. In the situation where the two lasers have different detunings ($\Delta_L\ne\Delta_P$), the four level system exhibits one single dark state $(|+\rangle-|-\rangle)/\sqrt{2}$, i.e. fully independent of the different laser parameters. In the stationary regime, the atom is shelved in this dark state and, therefore, is not scattering any photons.  As soon as the degeneracy is lifted, there is no dark state resulting in a finite fluorescence. Note that in the case that both lasers have the same detuning, one has the usual tripod configuration with two dark states. Therefore for the relevant energy level scheme of a barium ion, it is expected to show four MCPT states (two each for the D-state fine structure levels) independent of the relative laser detunings. In addition it is expected to have two more (one with each D-state fine structure levels) MCPT states between the ground and the D-states depending on the relative detuning of the lasers. Thus the system is rich in MCPT states which may be useful resource for quantum state manipulation having negligible interaction with its environment.

\section{Experimental setup}

The experiment has been performed on a single ion to avoid excess Doppler broadening and micro-motion while giving higher spatial resolution for magnetic field null point determination. However, for lower spatial resolution, a cloud of ions or an ion crystal may be
used.

\begin{figure}[htbp]
        \centering
        \includegraphics[width=.48\textwidth]
        {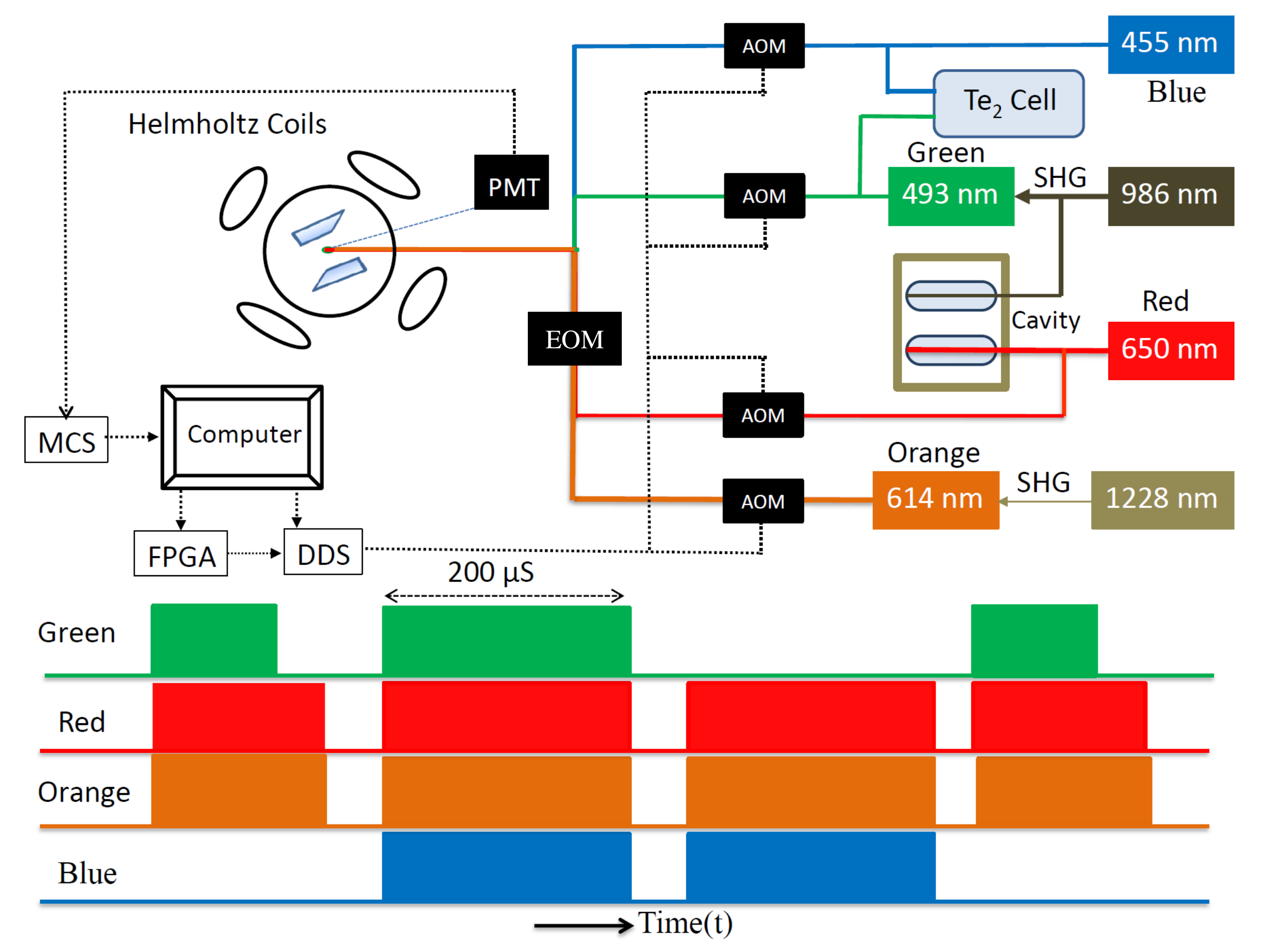}
    \caption[experimental setup]{A Schematic diagram of the experimental
setup. The laser frequency and amplitudes are
controlled by a acusto-optical modulators (AOM). All the laser
frequencies are monitored by a wavemeter with 5 MHz
resolution. A typical temporal pulse sequence generated by switching the AOMs via field-programmable gate arrays (FPGA) and direct digital synthesizers (DDS) for all the lasers used for
the measurement is also presented.}
    \label{fig:Total_setup}
\end{figure}

 A schematic of the experimental setup is shown in Fig.~(\ref{fig:Total_setup}),
further detail of the setup can be found in~\cite{Dem15,Dut16}. The
main components relevant to the present scheme are the ion trap, the
excitation lasers and the magnetic field coils. In the following a
brief description of these elements are given. The ion trap is a
linear Paul trap operating at a secular frequency of about $1$~MHz in
the radial while few hundreds kHz in the axial direction. The ion
temperature after Doppler cooling is about $0.5$~mK which is
sufficient for the present experiment as Doppler broadening does not
play a role. The ion fluorescence signal is detected in a photo
multiplier tube after it is filtered by a narrow band interference
filter with center wavelength at $493~$nm or alternatively at $455~$nm. Doppler
cooling is usually performed using $493$nm and $650$nm lasers
forming a closed cyclic transition. However as we will see in the later
section, we have alternatively used $455$nm, $650$nm and $614$nm
lasers together to laser cool the ion. The second approach leads to
a slightly higher temperature due to the higher width of the
P$_{3/2}$ level as compared to the P$_{1/2}$ level. All the lasers
are external cavity diode lasers (ECDL) with estimated linewidth of
about $500$~kHz. The $493$~nm laser is obtained by frequency doubling
of a $986$~nm ECDL inside a bow-tie cavity. The $614$~nm light on the
contrary is obtained by frequency doubling a $1228$~nm ECDL in a
single pass arrangement in a wave guide based periodically poled KTP crystal. In
this arrangement, the output power and frequency remains stable due to all fiber
based connectivity. In order to minimize the long term relative frequency drift of the
cooling lasers, $650$~nm laser is phase locked to a reference
cavity along with $986$~nm laser within a same zerodure spacer. The
$455$~nm laser is locked to a tellurium spectral line using
modulation transfer spectroscopy (MTS)~\cite{Dut16a}. The $493$~nm
laser can also be locked to the same tellurium setup if needed.
Magnetic fields are generated at the center of the trapping region
using three pairs of coils placed in an orthogonal setup as shown in
Fig.~(\ref{fig:Total_setup}) in Helmholtz configuration. This
arrangement ensures homogeneity of the field at the trap center. Our measurement scheme relies on the detection of fluorescence near zero magnetic field, which makes the Doppler cooling process inefficient due to the fact that ion goes out of the cooling cycle as it goes into the dark D-states. A polarization modulating electro-optic modulator(EOM) is used to modulate the polarization of repump lasers during the cooling pulse to break degeneracy in the D-states \cite{Berk04}.

A single aspheric lens with numerical aperture $0.4$ placed inside the vacuum chamber collects the fluorescence photons from the trapped ion. In order to suppress
the background photons from entering into the detection setup a narrow
band interference filter centered at $455~$nm is placed before a
photo-multiplier tube (PMT). In the case of $493~$nm photon detection, an interference filter with center wavelength at $493~$nm is used instead. The maximum photon count rate observed
is about $70,000$ count/sec on a background scattering rate of about
$10,000$ count/sec. The scattering is mainly from the metal
electrodes of the trap. As shown in Fig.~(\ref{fig:Total_setup}), the
photon detection setup is placed perpendicular to the excitation
laser beams thus observing only spontaneously emitted photons
without making any distinction of their polarization. Therefore the
configuration is similar to Hanle but not the measurements. \\

\section{Experimental procedure}

A single ion is loaded into the trap from a hot barium oven and
ionized resonantly using a $413$~nm ECDL laser at a loading
efficiency of about $1$ ion/min. Once a successful loading is done,
the ion usually remains in the trap for months. The ion is then
continuously excited by $455,~614,~650$~nm lasers which are combined
together in a fiber and injected into the trap. For the experiment
reported here the polarization of all the lasers are perpendicular to the magnetic field direction. However, the polarization of
the lasers do not play any role in the results which are reported
here. Two different measurements were performed by alternately detecting either the
$455~$nm photons or $493~$nm photons when ion is
continuously excited by $455,~614,~650$nm lasers and the $493~$nm laser is kept off respectively.\\

The saturation parameters of the lasers are experimentally obtained from
the fluorescence count measurement as a function of input laser
power while all other parameters remaining constant. In case of the
$455~$nm laser, the saturation parameter is also verified by
frequency shift measurement while probing the narrow quadrupole
transition between S$_{1/2}$ and D$_{5/2}$ states. The measured
saturation intensity obtained from these two methods agree within
their respective error of about $4.5\pm0.5~\mu$W with a focus spot diameter of about $100~\mu$m.

\section{Results}

\begin{figure}[htbp]
        \centering
        \includegraphics[width=.48\textwidth]
        {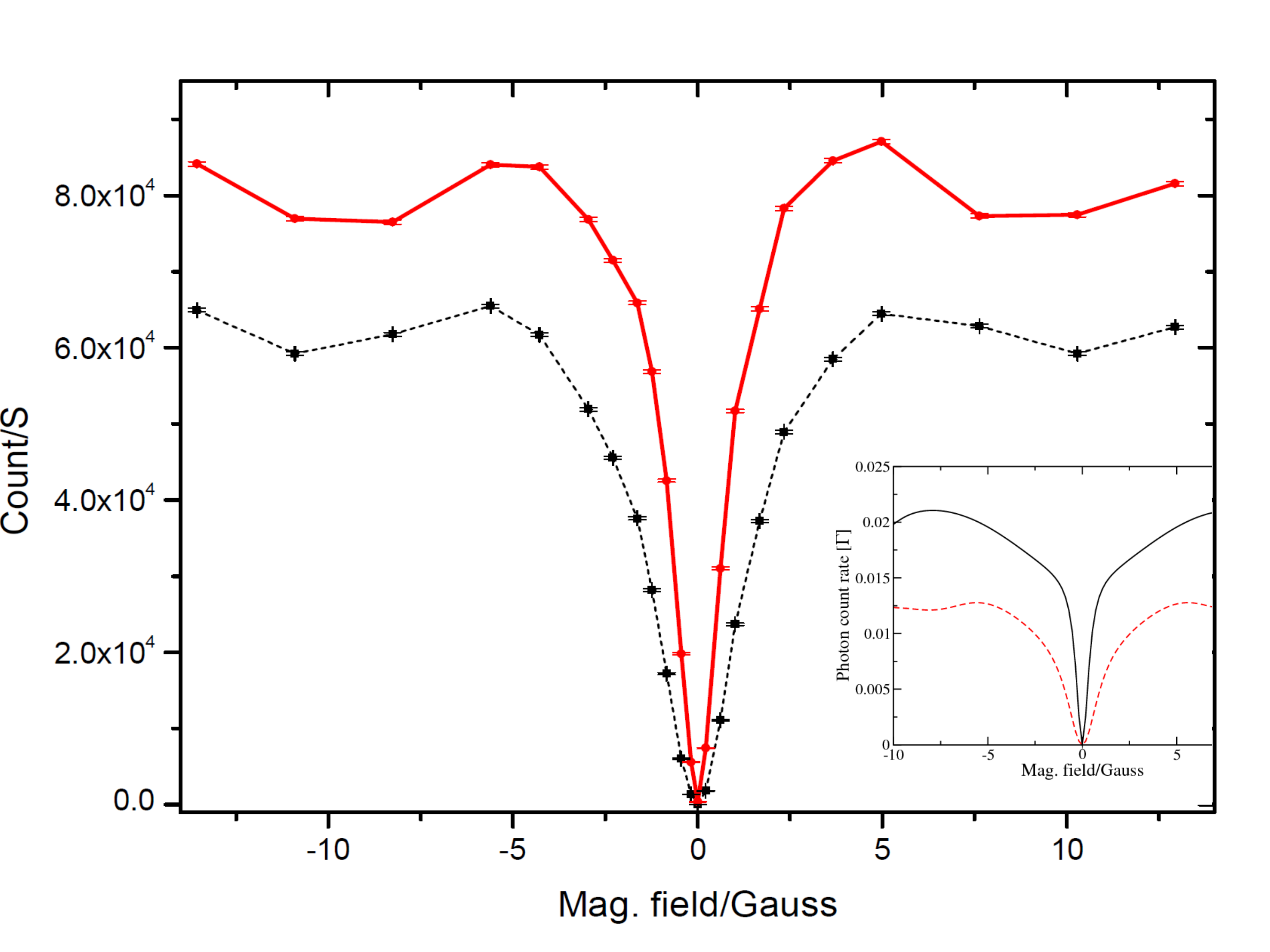}
    \caption[Fluorescence count]{The fluorescence at $455$nm as a function of the magnitude of the applied magnetic field. The dashed curve is obtained when $493$ nm
laser is also applied on the ion while the solid line is when it is
off. The right-side inset shows the fluorescence rate as a function of the magnetic field as calculated from theory using the same experimental parameters. The left-side inset shows the derivative of the dashed curve which determines the sensitivity of the rate of fluorescence as a function of magnetic field.}
    \label{fig:Fluorescence}
\end{figure}

In the following we discuss the results of the MCPT experiments. The narrowness of the MCPT feature is the figure-of-merit for the usefulness of this scheme. The resonance at zero magnetic field can be observed either by observing the spontaneous emission from P$_{1/2}$-S$_{1/2}$ (lower lambda) or by P$_{3/2}$-S$_{1/2}$ (upper lambda) transitions respectively. First we discuss the MCPT feature of the upper lambda as shown in Fig.~(\ref{fig:Fluorescence}). Then we show that the ultimate narrow feature possible in our system is obtained by using the lower lambda. The second approach allows measurement at zero field without any scattered photon background. We find that the width of this fluorescence dip is limited by the saturation parameter of the lasers and mainly the re-pump laser. The last result is the identification of a distinct bright state which energetically lies in close proximity to the dark state in the D-level Zeeman manifold. This state may be useful for readout of the dark state. \\
The magnetic field at the position of the ion can be precisely controlled by
applying current on individual pair of coils placed in an orthogonal
arrangement as shown in Fig.~(\ref{fig:Total_setup}). This allows independent control over the individual vector components of the field. The three vector components are
individually calibrated with respect to the coil current along that
component by performing spectroscopy on a narrow quadrupole transition between the
S$_{1/2}$-D$_{5/2}$ level corresponding to $\Delta$m$=0$. Therefore each magnetic field component along a coil direction is derived with a precision of $2.4\times 10^{-4}$ G.
In both the measurement schemes discussed below, the photons are detected while the $455$, $614$ and $650~$nm lasers are simultaneously and continuously applied as shown in Fig.~(\ref{fig:Total_setup}).


\begin{figure}[htbp]
        \centering
        \includegraphics[width=.48\textwidth]
        {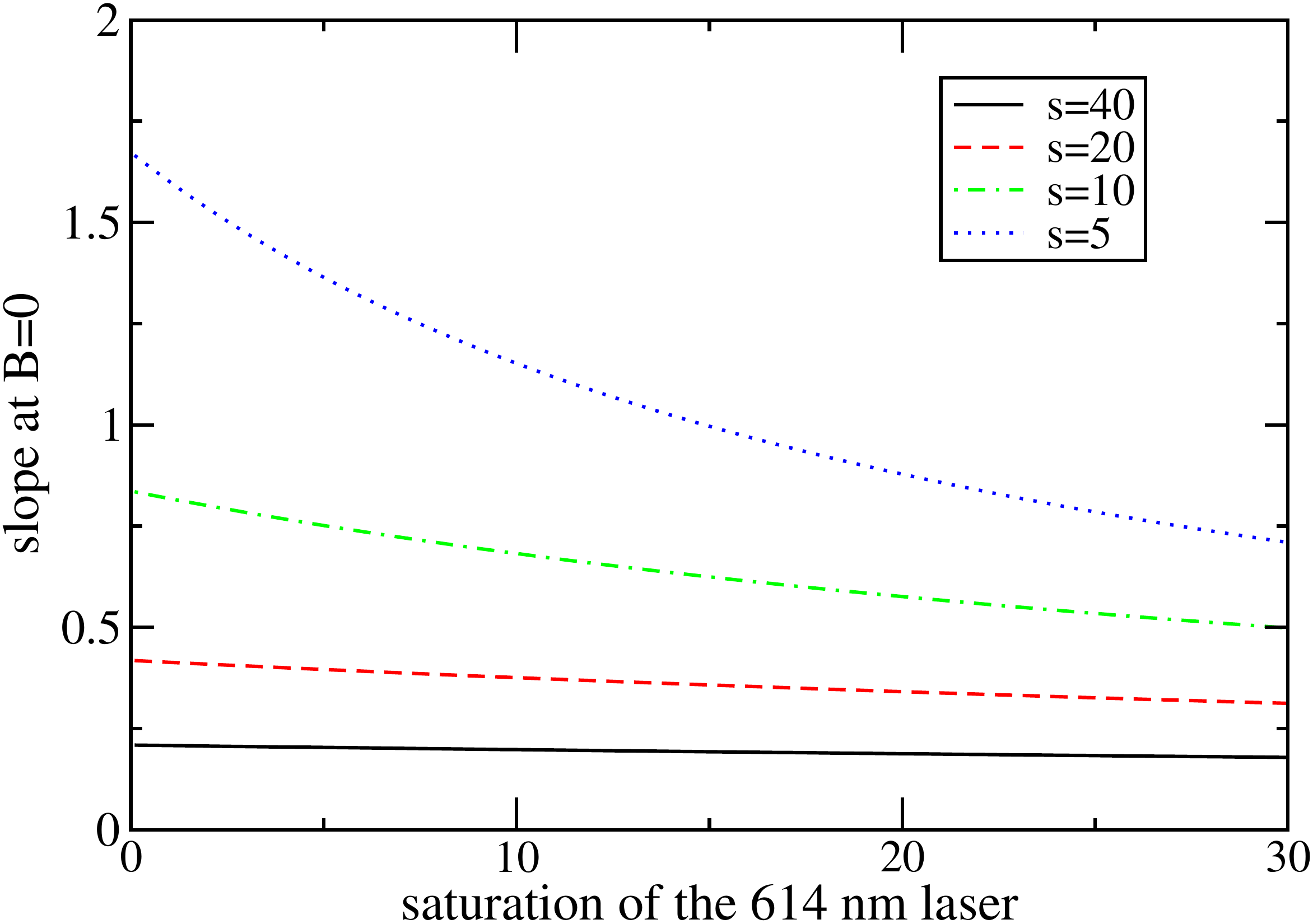}
    \caption{(Color online) Numerical value of the slope of the derivative of the fluorescence at $455$nm as a function of the saturation parameter of the $614$ nm laser and for different values of the saturation of the $650$ nm laser. As one can see, lowering the repumper intensities is increasing the slope of the dispersion curve and, thereby, the sensitivity of the experiment.}
    \label{fig:slope}
\end{figure}

The fluorescence count observed at $455~$nm as a function of the magnitude of the applied magnetic field is then plotted in Fig.~(\ref{fig:Fluorescence}). The dashed curve is obtained when $493~$nm laser is also applied on the ion while the solid line is when it is
off. The curve (not shown here) for higher magnetic field is rather very complex as it involves optical nutation due to coherence of the lasers interacting with $18$-level ion. However for our present discussion, these features do not play a role. The right inset in fig.~(\ref{fig:Fluorescence})shows the numeric calculation of the same data for parameters that are used in the experiment. Even though the numeric result matches well for the central feature it is not so good in describing the features for higher magnetic field in which case the sensitivity to small changes in intensity or frequency is high. Nevertheless, we have found a qualitative agreement with the experimental results as shown in Fig.~(\ref{fig:Fluorescence}), especially in the case when the 493 nm laser is on. The numerical values for the different laser parameters used in the experiment are: saturation of the $455$ nm (resp. 614, 650 and 493) is 0.5 (resp. 15, 40 and 5 (when on)) and the detuning of the $455$ nm (resp. 614, 650 and 493) is -10MHz (resp. -50MHz, -40MHz and -20MHz). The width in this scheme is dominated by the intensities of the re-pumps and the zero field fluorescence count is determined by the scattered background of the $455~$nm photons. The good agreement with the theory even for this multi-level system led us to study the behavior of the sensitivity as a function of different experimental parameters. Indeed, from the numerical computations as shown in Fig.~(\ref{fig:slope}), it turns out that the shape of the dip is primarily independent of the different laser detunings and only dependent of the  saturation parameters of the 614nm and 650nm (re-pump) laser, whereas the wings, i.e. the large magnetic field behavior, depend much more on all the lasers parameters.\\

\begin{figure}[htbp]
        \centering
        \includegraphics[width=.52\textwidth]
        {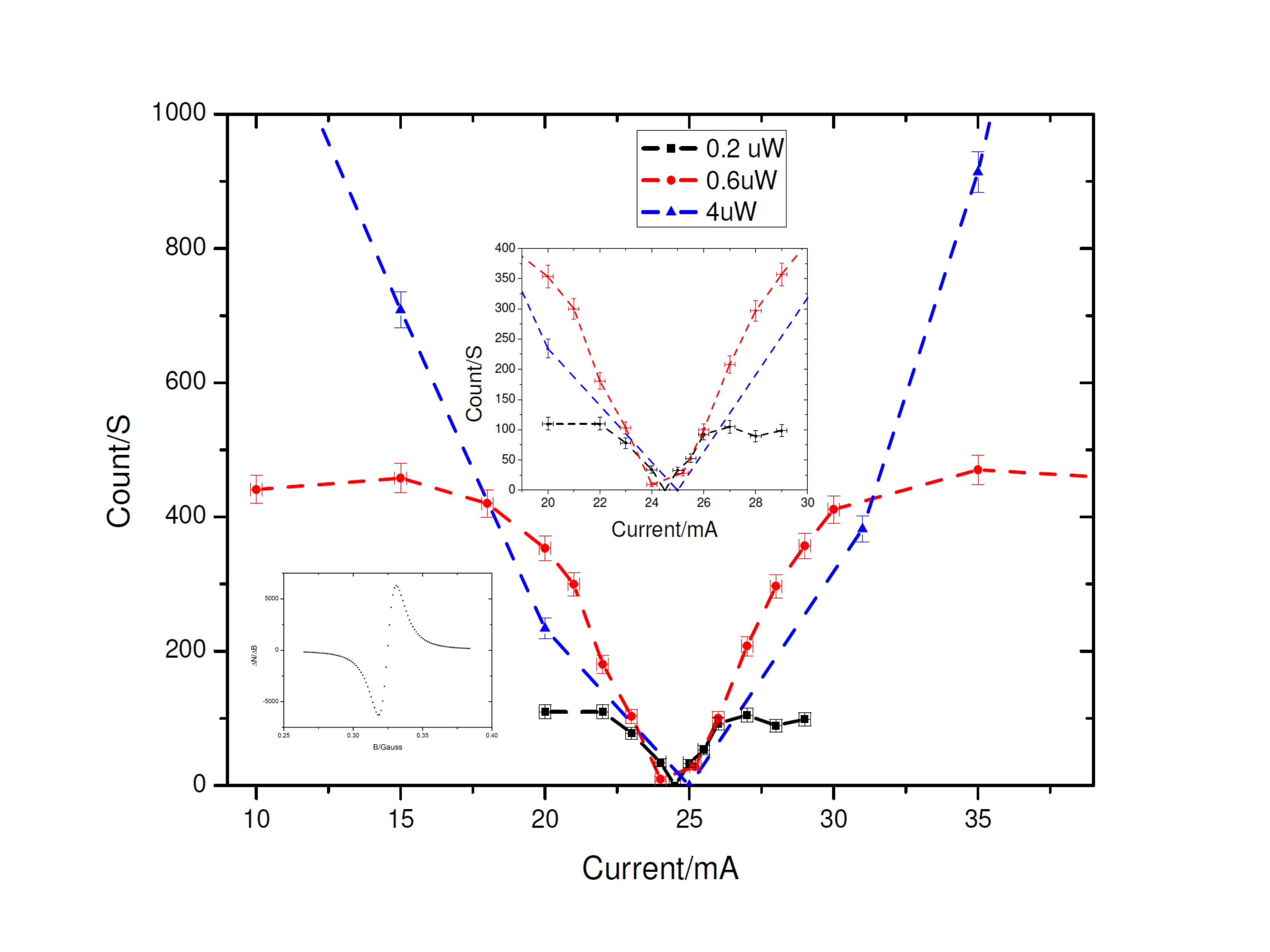}
    \caption[Fluorescence count]{The fluorescence at $493$nm as a function of the magnetic field
    strength for three different saturation parameter of the re-pumps. The zoomed portion of the
    graph is shown in the inset above while derivative of the black curve with respect to magnetic
    field is plotted on the inset to the left. The slope at the zero magnetic field obtained is about $1.61\times10^{-6}$ counts/s/gauss.}
    \label{fig:plot}
\end{figure}

The same experiment is now repeated with the background free detection scheme. In this case the lower lambda systems is used and the fluorescence at $493~$nm is plotted as a function of the magnetic field strength for different re-pump saturation parameters. The width of the central narrow dip as shown in Fig.~(\ref{fig:plot}) is observed to be reduced as the saturation parameters is decreased at the same time the total count also goes down as is expected from fig.~(\ref{fig:slope}). However when the saturation of these lasers are too low and the magnetic field is non-zero most of the population in steady remains in the D-states and hence the photon counts reduce. Thus the integration time becomes longer.

In both the above schemes at the exact null magnetic field, the atom is transparent to all the
laser lights and no fluorescence is observed except only background
scattering level. In case of the second experiment the background is only the dark count of the detectors since there is no excitation laser at $493$~nm applied. As explained by the simple model, the atom goes to one of the dark states as soon as the magnetic field is null which makes the magnetic sub-levels degenerate. Since the D-states are
only coupled by quadrupole transitions, the width of this dip is
mainly governed by the saturation parameter of the re-pump lasers as shown
in Fig.~(\ref{fig:slope}). The sensitivity which is plotted here, ultimately governs the sensitivity of our scheme when applied to magnetic field detection or for laser locking.\\


Now we discuss the three major differences in the spectral feature
in our experiment as compared to a gas cell experiment. First,
unlike a gas cell experiment with the Hanle type CPT induced by zero
magnetic field, our experiment with a single atom does not show any
shift due to the longitudinal or transverse magnetic field. It is
completely symmetric and hence only three axis coil current scan to
nullify the magnetic field in three orthogonal direction allows the
measurement of the vector field. In a gas cell the shift of the
resonance center in presence of a longitudinal magnetic field along
the light propagation direction originates due to the velocity
distribution of the free atoms. In a single ion experiment confined
in a trap, cooled to the Doppler limit, there is no net shift in the
spectrum. Second, in our experiment, there are three possible CPT states
depending on the relative laser detunings. Therefore in one
configuration, namely lower lambda, it is possible to achieve detector shot noise limited sensitivity as shown in Fig.~(\ref{fig:plot}). In case of a gas cell experiment with the best possible scheme, the sensitivity is limited by the background light scattering which is usually removed by a lock-in type measurement. Last but not the least, the measurement volume of a single ion is determined by the single ion confinement volume which under Doppler
limit is a few hundred micrometers cube. This is orders of magnitude
lower as compared to small gas cells and hence can provide high
spatial resolution. One essential point to note is the possibility to measure the null field in our setup which is not possible in other sensitive magnetic field sensors like SQUID \cite{Jak64} due to the requirement of a bias field.\\

\begin{figure}[htbp]
        \centering
        \includegraphics[width=.48\textwidth]
        {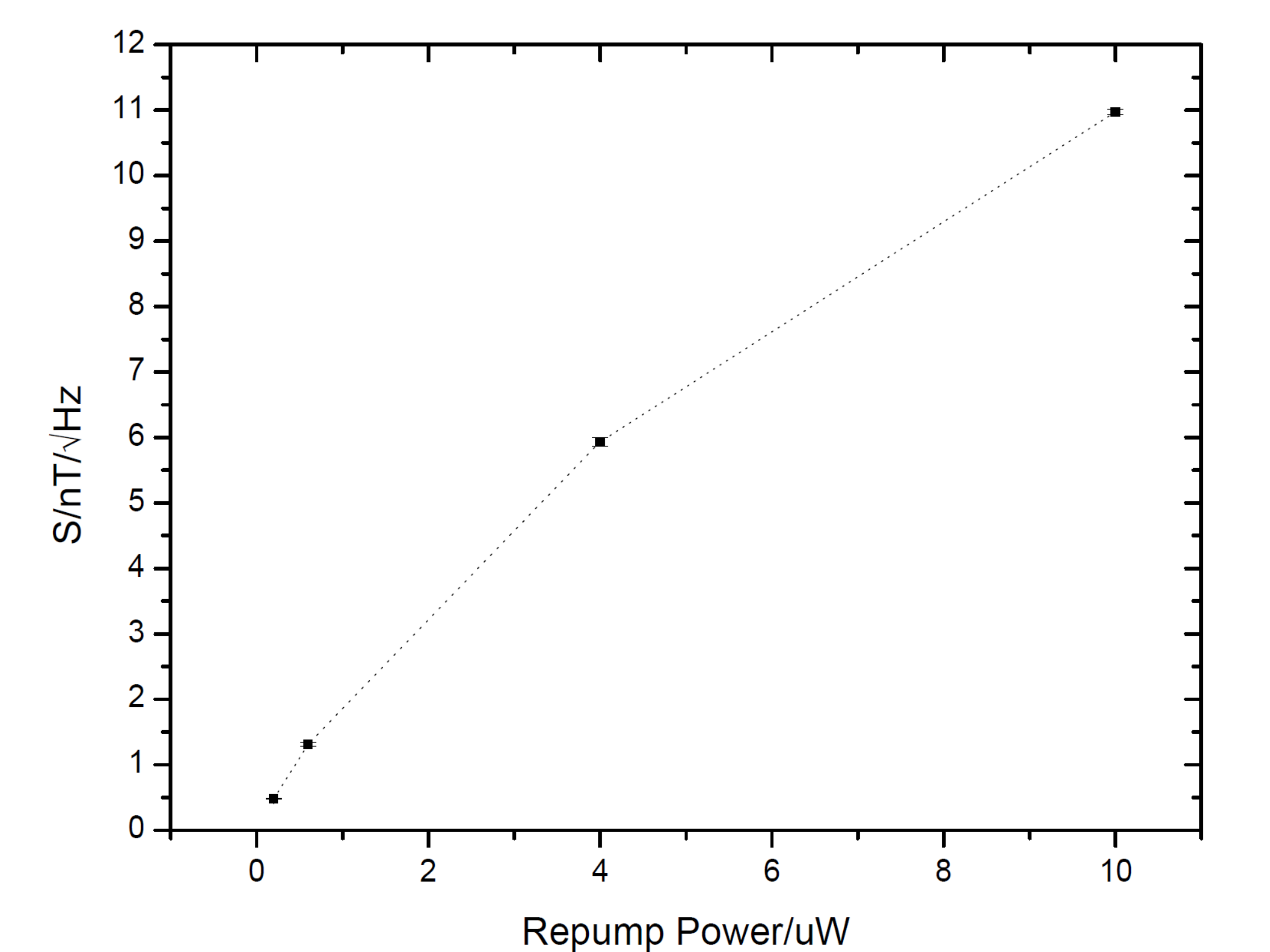}
    \caption[Sensitivity vs repump powers.]{Measured sensitivity as function of the re-pump laser power for the background free detection of $493~$nm photons. Both the re-pump lasers are kept at same power during a measurement. It was not possible to detect any signal below a power of $200$nW.}
    \label{fig:sensitivity}
\end{figure}

Now we provide some of the key figures-of-merit for our present
system which is pertinent to its usefulness.  The sensitivity of field measurement is dependent on the derivatives of the narrow dip which is $\Delta N/\Delta B \approx 1.61\times10^6$~counts/s/Gauss. The present detector background noise $(\sigma (N))^2$ is about $60$ counts/s. Thus the sensitivity which is defined as
\begin{equation}
\Delta B = \sigma (N)/(\Delta N/\Delta B),
\end{equation}
\begin{figure}[htbp]
        \includegraphics[width=.48\textwidth]
        {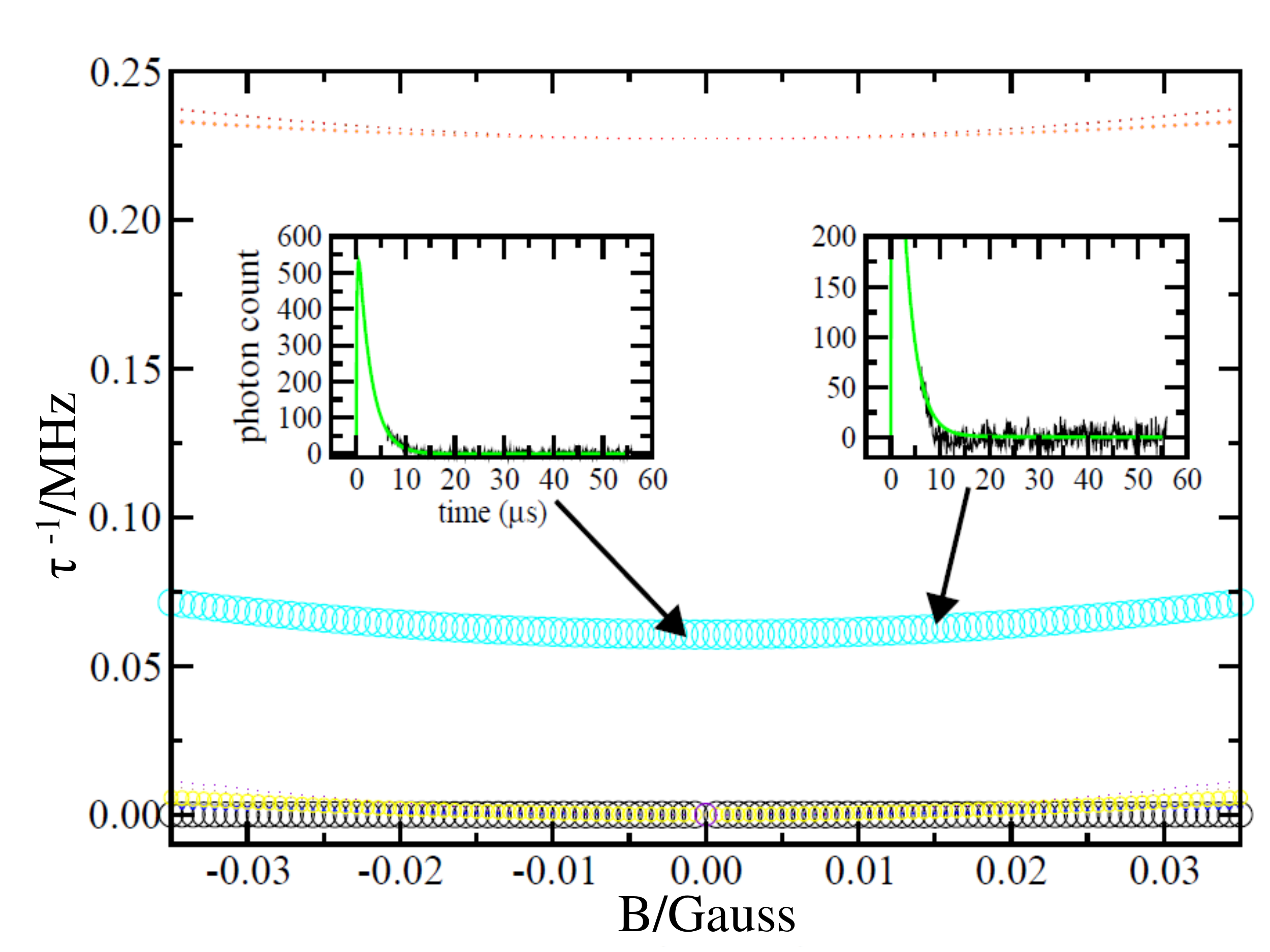}
    \caption[Dark and bright states]{Decay rate of numerically obtained intermediate states are plotted as a function of applied magnetic field. The size of
    the circles is proportional to the overlap with the initial density matrix. The two inset shows the fluoresces decay rate for two specific magnetic fields as example. The data fits well with the numerical decay rate (light green) for the same parameters as used in the experiment.}
    \label{fig:DarkBright}
\end{figure}
is about $470~$pT$/\sqrt{\text{Hz}}$. Fig.~(\ref{fig:sensitivity})
shows the measured sensitivity as obtained from
Fig.~(\ref{fig:plot}) for different re-pump intensities. This
sensitively is not at par with other schemes for magnetometry namely
single ion \cite{All02,Kom03,Bau16,Aleksandrov,Budker1} or SQUID
\cite{Clarke} etc, however it is an alternative approach to measure
the field. More important this technique allows to lock to null
magnetic field at the ion position which is important in some
experiments. Moreover the narrow MCPT feature may be useful to lock
laser frequency as shown recently in~\cite{Winc17}. Moreover if one
wants to use the dark states for quantum state manipulation due to
their insensitivity to external fields, it will be necessary to find
a close lying bright state to read out the dark state.
Therefore in the following we looked at the time evolution of the dark state formation as it attends steady state. \\

The temporal evolution of the MCPT state is observed by measuring the time evolution of the photon counts
from the P$_{1/2}$ state after the ion is excited into the P$_{3/2}$ state.
The population in the P$_{1/2}$ evolves such that the final state is one of the MCPT states.
In the time scale of experiment, it turns out there is only one bright state with significantly
long life-time (of about $10\mu$s). Fig.~(\ref{fig:DarkBright}) shows the theoretical lifetime of the relevant eigenstates
of the Liouville operator as a function of applied magnetic field. As observed in the experiment close to the null magnetic
field the P$_{1/2}$ state decays with an exponent which agrees well with that predicted by theory.
The exponential decay as well as the theoretical fit for the experimental parameters used is shown for
two values of the magnetic field. A closer look at the structure of the ``density matrix''  associated
to this bright state reveals that it is ``mixed state'', i.e. a linear superposition of density matrices of pure states,
but with positive and negative weights (being a decaying state, its trace has to vanish). The pure states actually
qualitatively correspond to the dressed states of the system in presence of the three lasers,
which includes the four dark states. The weights  ``density matrix'' are such that it
is mainly leaving in the D$_{3/2}$ manifold (50\% of the population) and its lifetime is then due to the coupling to the
P$_{1/2}$ manifold (without the laser at 650~nm, it would become another dark state). Finally, at $B=0$, this decaying state
is not coupled to the dark state manifold when applying a magnetic field but only to other bright states, which explains that
its decay rate increases quadratically with $B$.
All these features can be qualitatively understood with the simple model shown in Fig.~\ref{fig:simple_model}.
As explained above, for $\delta=0$, the steady-state is the dark state
$\propto |+\rangle-|-\rangle$. The bright state with the smallest decay rate corresponds to a mixed state involving
the four possible dressed states. In the parameter regime comparable to the experiment, it turns out that
three of the dressed states are qualitatively corresponding to $g\rangle$, $|e\rangle$, $|+\rangle+|-\rangle$ and the
fourth one is the dark state. In addition, the population in the excited $|e\rangle$ is negligible, explaining a
decay rate few orders lower than $\Gamma$.

This peculiar structure of this bright state, in particular the large weight in the dark state manifold,
may play a significant role in using the dark state for quantum information processing while reading
out the state by coupling to the bright one.

\section{Conclusion and outlook}

We conclude that a double lambda system may be used to obtain narrow spectral feature dependent on applied external magnetic field near null value. The background free experiment reaches a sensitivity of $470~$pT/$\sqrt{\text{Hz}}$ to the change in magnetic field. Our experimental result further confirms the presence of an unique bright state with significant lifetime near the dark state in the D-level Zeeman manifold. This may be further explored for dark state readout in quantum information processing.


\end{document}